\documentclass[preprint]{aastex62}
\usepackage{rotating}
\submitjournal{ApJ Letters}

\shorttitle{Venusian Tides}
\shortauthors{Green et al.}

\begin{document}

\title{Consequences of Tidal Dissipation in a Putative Venusian Ocean}

\correspondingauthor{Mattias Green}
\email{m.green@bangor.ac.uk}

\author{J. A. Mattias Green}
\affil{School of Ocean Sciences, Bangor University \\
 Menai Bridge, LL59 5AB, UK}

\author{Michael J. Way}
 \affiliation{NASA Goddard Institute for Space Studies, New York, New York, USA}
 \affiliation{GSFC Sellers Exoplanet Environments Collaboration}
 \affiliation{Theoretical Astrophysics, Department of Physics and Astronomy, Uppsala University, Uppsala, SE-75120, Sweden}

\author{Rory Barnes}
\affiliation{Department of Astronomy, University of Washington, Seattle, Washington, USA}

\begin{abstract}
The solar tide in an ancient Venusian ocean is simulated using a dedicated numerical tidal model. Simulations with varying ocean depth and rotational periods ranging from -243 to 64 sidereal Earth days are used to calculate the tidal dissipation rates and associated tidal torque. The results show that the tidal dissipation could have varied by more than 5 orders of magnitude, from 0.001--780 Gigawatts, depending on rotational period and ocean depth. The associated tidal torque is about 2 orders of magnitude below the present day Venusian atmospheric torque, and could change the Venusian day length by up to 72 days per million years depending on rotation rate. Consequently, an ocean tide on ancient Venus could have had significant effects on the rotational history of the planet. These calculations have implications for the rotational periods of similarly close-in exoplanetary worlds and the location of the inner edge of the liquid water habitable zone.
\end{abstract}


\section{Introduction}
It has been argued that Venus may have had an ocean in its deep past \citep{Hashimoto2009,Hamano2013,Shellnutt2019}, and hence it may have been habitable if its rotation rate was similar to today's \citep{Way2016}. An ocean also raises the prospect of a solar ocean tide, and an associated tidal drag that could have affected the rotation. Here, we explore the subject of a Venusian ocean further by investigating tidally driven dissipation rates on ancient Venus to understand and constrain its history. This can also help inform studies of ocean-bearing exoplanets where the rotation rate is critical to understanding climate dynamics \citep[e.g.,][]{Yang2014,Way2018}.

There are several reasons why an understanding of tidally driven processes on other planets is important. Ocean tides on Earth are a key driver of the evolution of the orbital configuration of the system through tidal friction \citep{munk68, billsray99, Green2017}, and they have a profound impact on Earth by providing some of the energy that powers vertical fluxes of carbon and nutrients \citep[e.g.,][]{sharplesetal07}, and sustaining deep water formation at high latitudes by driving vertical volume fluxes through mixing \citep{munk66, munkwunsch98}. Tides have also been recognized as a potential driver for evolution and mass extinction events \citep{Balbus2014}. These effects could be much stronger on other worlds \citep[e.g.,][]{Barnes2013}, and a broad understanding of tidal dissipation over a range of planetary and orbital parameters could help our understanding of planetary evolution, as well as guide our search for life beyond Earth. It makes sense to start such simulations for a well-studied planet with an observed topography, rather than more speculative simulations for other exoplanets.

In this paper we aim to describe the plausible range of tidal dissipation rates in an ancient Venusian ocean, and the associated effect the tide may have on the planet's rotation. We start by describing our dedicated tidal model in the next section, and follow up with the results in Section 3 and a summary in section 4.

\section{Methods}
\subsection{Tidal modelling}
The Venusian tides were simulated using the portable Oregon State University Tidal Inversion Software (OTIS), which has been used extensively to simulate deep-time, present day, and future tides on Earth \citep[e.g.,][]{Egbert2004, Wilmes2017, Green2017, Green2018}. OTIS provides a numerical solution to the linearized shallow water equations, with the non-linear advection and horizontal diffusion excluded without a loss in accuracy \citep{Egbert2004}:
%
\begin{eqnarray}
  \frac{\partial \mathbf{U}}{\partial t} + f\times\mathbf{U} &=& -gH\nabla(\eta - \eta_{SAL} - \eta_{EQ}) - \mathbf{F} \\
   \frac{\partial\eta}{\partial t} - \nabla\cdot\mathbf{U}&=& 0
\end{eqnarray}
%
Here, $\mathbf{U}=\mathbf{u}H$ is the depth-integrated volume transport ($\mathbf{u}$ is the horizontal velocity vector and $H$ is the water depth), $f$ is the Coriolis parameter, $g$ is acceleration due to gravity, $\eta$ is the sea-surface elevation, $\eta_{SAL}$ is the self-attraction and loading elevation, $\eta_{EQ}$ is the elevation of the equilibrium tide, and $\mathbf{F}$ the tidal dissipation term. $\mathbf{F}$ can be split into two parts, $\mathbf{F} = \mathbf{F_B + F_w}$. Here, $\mathbf{F_B}$ simulates bed friction between the liquid ocean and the solid planet, and $\mathbf{F_w}$ represents energy losses due to tidal conversion, i.e., the generation of a baroclinic or internal tide within a stratified water column \citep[see][for an introduction]{garrett03}. Bed friction is parameterized through the standard quadratic law: $\mathbf{F_B} = C_d\mathbf{u|u|}$, where $C_d$=0.003 is a dimensionless drag coefficient. The chosen value for the drag is the standard bed roughness for Earth and is determined by the roughness of the seafloor. Two simulations were performed where $C_d$ was set to 0.009 or 0.001 (not shown), and they did not significantly change the results. The tidal conversion term, $\mathbf{F_w}$ can be written as $\mathbf{F_w} = C\mathbf{U}$. The conversion coefficient, $C$, was computed following \citet{Zaron2006} \citetext{see \citealp{greennycander13} and \citealp{Green2013a} for details}:
%
\begin{equation}\label{eq:C}
  C(x,y)=\gamma \frac{N_H\bar{N}(\nabla H)^2}{8\pi\omega}
\end{equation}
%
Here, $\gamma=100$ represents a dimensionless scaling factor representing unresolved bathymetric roughness, $N_H$ is the buoyancy frequency at the seabed, $\bar{N}$ represents the vertical average of the buoyancy frequency, and $\omega$ is the frequency of the tide. The buoyancy frequency, $N$, used to compute $N_H$ and $\bar{N}$ is defined as $N^2 = -g/\rho \partial\rho/\partial z$, but is unknown for an ancient Venusian ocean. Consequently, we used one based on a statistical fit of that observed on present day Earth: $N(x,y)=0.00524\exp(-z/1300)$, where $z$ is the vertical coordinate, and the constants 0.00524 and 1300 have units of $\mathrm{s^{-1}}$ and $\mathrm{m}$, respectively \citep{Zaron2006}. To test robustness of this choice we did simulations for all our rotation rates without any tidal conversion by setting $\gamma=0$ in Eq.~(\ref{eq:C}), representing an unstratified ocean and denoted ``noIT'' for ``no Internal Tides'' in the following. For a few scenarios (see Table~\ref{tab:runs}) we performed sensitivity simulations where $\gamma$ was increased by a factor 10 (denoted ``ITx10'' in the following) to simulate a very strongly stratified ocean. These two extreme cases will act to provide a very wide sensitivity range; see \citet{Green2017} for a case study on Earth.

\subsection{Forcing and boundary conditions}
\begin{table}[!t]
  \centering
  \caption{Parameters used in the initial model simulations; see \citet{Way2016} for details.}\label{tab:params}
  \begin{tabular}{l|l|l}
     \hline
     Parameter & symbol	& Value\\
     \hline
Gravity & $g$ &	8.87 $\mathrm{m\ s^{-2}}$ 	\\
Radius & $R_v$ &	6052 km  	\\
Rotation period & $T_{sidereal}$ &	-243.025 Earth days	 	\\
Year length & $T_{orbit}$ &	224.701 Earth days	 	\\
Solar distance & $r$ &	$\mathrm{108.2\times 10^9}$ m 	\\
Solar day & $T_{sol}$	& 116.75 Earth days 	\\
Tidal period & $T_{S2}$ &	58.375 Earth days 	\\	
Mass of Venus & $m_v$ &	0.815$m_E$, $\mathrm{4.867\times 10^{24}}$ kg  \\
Mass of Sun & $m_S$ &	332946$m_E$, $\mathrm{1.989\times 10^{30}}$ kg  \\
     \hline
  \end{tabular}
\end{table}
The tide on Venus will be dominated by a semi-diurnal solar tide. Because of Venus' small obliquity and eccentricity, diurnal tides can be neglected \citep{hendershott77}. Even if Venus had a large obliquity and/or eccentricity in the past, it is ignored at this stage as it would add yet another uncertainty. The equilibrium tidal elevation and frequency of the solar tide in the model was consequently set to represent conditions on Venus. The equilibrium solar tide is directly proportional to the mass of the Sun, and inversely proportional to the cube of the distance between the planet and the Sun. Consequently, the solar equilibrium tide on Venus is 2.67 times larger than on Earth (see Table~\ref{tab:params} for the numerical values used).

80--85\% of Venus has been resurfaced over the past several hundred million years \citep{Kreslavsky2015, IvanovHead18} and the bathymetry of an ancient Venusian ocean is thus unknown. There is, however, modern topography available from the Venus Magellan mission, and we used that as one proxy (the other being modern Earth's ocean topography) for the past topography \citep[available from http://pds-geo sciences.wustl.edu/mgn/mgn-v-rss-5-gravity-l2-v1/mg\_5201; see][for details]{FordPettengill1992}. The vertical resolution of the data is approximately 80 m, and the  horizontal resolution, which was also used in our model simulations, was $\mathrm{1^\circ\times 1^\circ}$ \citep{FordPettengill1992}.

\begin{figure}
\includegraphics[width=\textwidth,trim={0 3cm 0 12cm},clip]{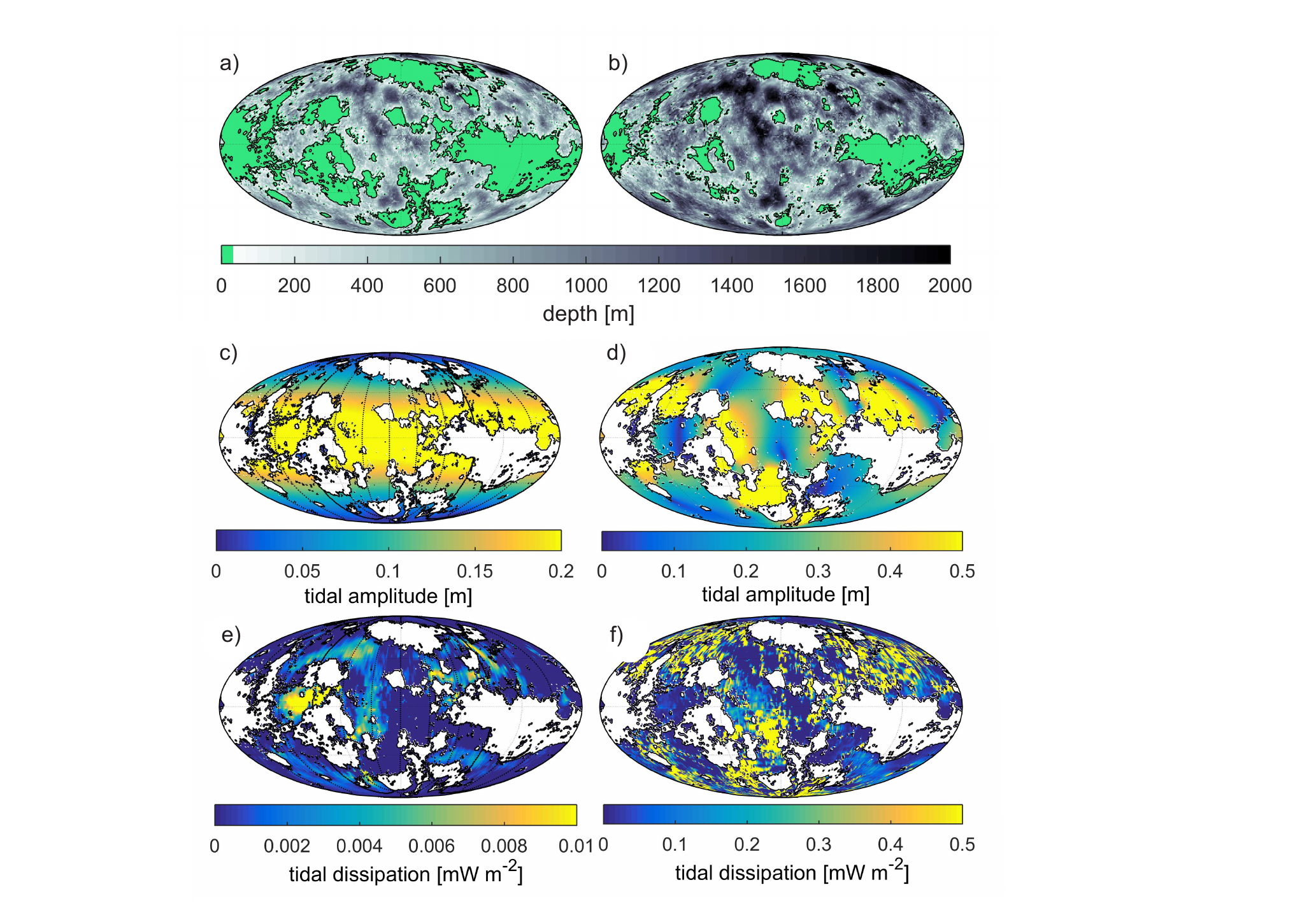}
\caption{a)--b) The ocean bathymetry (depth in m; land is green) for the two configurations. a) shallow; b) deep. Note that the depth scale saturates in the deep oceans -- in the shallow simulation the deepest point is 2340m.
\newline
c) Solar tidal amplitudes (in meters) in a present day ocean on Venus from the shallow simulation. \newline
  d) As in c) but for the simulation with an 8-earth day prograde rotation rate. \emph{Note the different color scales between panels.}\newline
 e--f) As in panels c--d, but showing the tidal dissipation rate in $\mathrm{mW\ m^{-2}}$.}\label{fig:bathy}
\end{figure}
Runs were completed for two different depth configurations: shallow and deep. For the shallow runs, any land in the bathymetry below the mean radius of Venus \citep[6051.84km, see][]{FordPettengill1992} was set as ocean, whereas all of that above the mean radius was set to land. This gave an ocean with a mean depth of 330 m, similar to that used in the work of \cite{Way2016}. The deep simulations had 500 m of water added to the shallow bathymetry, leading to an average depth of 830 m and an average pressure at the ocean floor similar to the atmospheric pressure of Venus today (note that there are no atmospheric effects included in our tidal model). In the shallow bathymetry, 69\% of Venus's surface area is ocean, whereas in the deep bathymetry this value increases to 80\% (see Fig.~\ref{fig:bathy}).

\subsection{Simulations and validation}
\begin{table}[!t]
\scriptsize
 \centering
  \caption{Summary of the simulation details. Note that all simulations were initially done with the shallow bathymetry, and repeated without any tidal conversion and are denoted in the following with ``no IT'' appended to the simulation name. Similarly, runs done with an ocean where 500 m had been added to the shallow bathymetry are denoted ``deep'' below (see the text for details).}\label{tab:runs}
 \begin{tabular}{|l|l|r|r|}
  \hline
	 & 		 & 	Solar day 	 & 	sidereal day 	\\
Simulation	 & 	Note	 & 	[Earth days]	 & 	 [Earth days]	\\
  \hline
Earth	 & 	Earth control	 & 	1	 & 	0.9972	\\
Venus & Present Day Venus orbit and shallow bathymetry & -116.75 & -243\\							
Venus ITx10 & $\gamma = 1000$& -116.75 & -243 \\							
Venus -05 & Present Day Venus daylength halved & -78.86 & -121.5\\							
Venus -64 & Earth's daylength x16, retrograde,& -50.66 & -64\\							
Venus -16  & Earth's daylength x16, retrograde,& -15.01 & -16\\							
Venus -8  & Earth's daylength x8, retrograde,& -7.725 & -8\\							
Venus -1  & Earth's daylength, retrograde,& -0.9956 & -1\\							
Venus -1 ITx10  & $\gamma = 1000$, Earth's daylength, retrograde,& -0.9956 & -1\\							
Venus 1 & Earth's daylength x1, prograde & 1 & 0.9972\\							
Venus 1 ITx10 & $\gamma=1000$, Earth's daylength x1, prograde & 1 & 0.9972\\							
Venus 8 & Earth's daylength x8, prograde & 8.31 & 8.02\\							
Venus 12 & Earth's daylength x12, prograde & 12.67 & 12\\							
Venus 16 & Earth's daylength x16, prograde & 17.27 & 16.04\\							
Venus 32 & Earth's daylength x32, prograde & 37.31 & 32\\							
Venus 64 & Earths daylength x64, prograde & 89.78 & 64.15\\							
Venus 64 ITx10 & $\gamma=1000$, Earths daylength x64, prograde & 89.78 & 64.15\\							
 \hline							
 \end{tabular}
\end{table}
The initial set of simulations were for both depths (330m \& 830m) for present day Venus' rotational parameter space, and repeated with ``no IT'' conditions (see methods for details). It has been suggested that Venus may have been rotating faster in the past, and possibly prograde \citep[e.g.][]{goldsoter79, dobrovolskisingersoll80, dobrovolskis80, CorreiaLaskar2001, CorreiaLaskar03, Correiaetal2003}. Furthermore, exoplanets could have a wide range of rotational periods, so we performed a series of sensitivity simulations over a range of rotation rates. The first had a day that was half of the present, or -121.512 days; this is called simulation 05 in Table \ref{tab:runs}. It was followed by simulations with retrograde rotation periods of -16, -8, and -1 days. We then extended the simulation set into prograde rotations with daylengths of 1, 8, 12, 16, 32, and 64 Earth days; these simulations are henceforth referred to by their respective rates.

The associated period of the solar tide is equal to half the solar day, where the latter is given by
%
\begin{equation}\label{eq:Tsol}
  \frac{1}{T_{sol}} = \frac{1}{T_{sidereal}}-\frac{1}{T_{orbit}}
\end{equation}
%
Note that $T_{orbit}<0$ for retrograde motions. See Table~\ref{tab:runs} for details about the simulations.

Each simulation lasted 20 tidal periods; 7 periods were used for harmonic analysis of the tide after a 13-period spin-up. A sensitivity test for the shallow simulation (not shown) was done when the simulation time was doubled, and there was no discernible difference between the simulations, i.e., the model converged. The model output consists of the amplitudes and phases of the sea-surface elevation ($\eta$) and the transports ($\mathbf{U}$).

Present day simulations (not shown) of the solar ($\mathrm{S_2}$) tide on Earth at 1$^\circ$ resolution has a root mean square error of 10 cm when compared to the altimetry constrained TPXO9-solution \citep[][]{Egbert2002}\footnote{http://volkov.oce.orst.edu/tides/global.html}. The associated globally integrated dissipation rate on Earth is overestimated by a factor 2 in this simulation. We can therefore expect our Venusian dissipation rates to be overestimates because of the lack of resolution of the bathymetry.

\subsection{Computations}
The tidal dissipation rates, $D$, were computed using the model output following
\citet{egbertray01} as the difference between the work $W$ done by the tide-generating force and the divergence of the energy flux $P$, i.e.,
%
\begin{equation}\label{eq:d}
  D = W-\nabla\cdot P
\end{equation}
%
with
%
\begin{eqnarray}
  W &=& g\rho \langle\mathbf{U}\cdot\nabla(\eta_{SAL}+\eta_{EQ})\rangle \\
  P &=& g\rho \langle\eta\mathbf{U}\rangle
\end{eqnarray}
%
where the angular brackets mark time-averages over a tidal period.

The associated tidal torque, $\tau$, can be written as:
%
\begin{equation}\label{eq:torque}
  \tau = \frac{3}{2}\frac{kGm_S^2R_v^5}{r^6}\sin{2\alpha}
\end{equation}
%
Here, $G$ is the gravitational constant, $m_S$ is the Sun's mass, $R_v$ is Venus' radius, $r$ is the Venus-Sun distance, and $k$ is a Love number that takes the non-uniformity of the planet into account. Because most of the bulge is assumed to be made of sea water we use $k=0.19$ as this is close to the ratio between Venus' average mass density and that of water, or 1/5.24 \citep[see][for details]{Macdonald64}. $\alpha$ is the lag angle between the tidal bulge and the planet-satellite axis; on Earth today $\sin(2\alpha)\sim 1/13$ \citep{Macdonald64}. We compute $\sin{2\alpha}$ for each simulation as the ratio between the tidal dissipation and the work done by the tide-generating force, $D/W$. This also allows us to compute the tidal damping factor (the number of cycles to obtain an e-folding decay of the amplitude) defined as $Q=W/D=1/\sin(2\alpha)$.

Calculation of the torque from Eq.~(\ref{eq:torque}) now allows us to calculate the resulting spin-down of the planet's rotation $\Omega_v$ from
%
\begin{equation}\label{eq:dOdt}
 \frac{d\Omega_v}{dt} = \frac{15}{4}\frac{kGm_s^2R_v^3}{m_vr^6}\sin{2\alpha}
\end{equation}
%
where $m_V$ is the mass of Venus.

\section{Results}
\subsection{Shallow results}
The shallow simulation shows that Venus would host only very small tides -- a few cm above the equilibrium tide -- if it had an ocean today (Fig.~\ref{fig:bathy}c). Consequently, the dissipation rates are very small, and measured in fractions of $\mathrm{mW\ m^{-2}}$ (Fig.~\ref{fig:bathy}e). The horizontally integrated rate in the shallow simulation is a mere 0.15 Gigawatts (GW) (see Fig.~\ref{fig:DvsR}a, which is discussed in detail below). This is a fraction of the dissipation of 600 GW from the solar tide in Earth's oceans today. The results make sense dynamically, because a significant amplification of the tide can only occur if the natural resonant period of an ocean basin is close to the tidal period, as is the case of the present day North Atlantic on Earth \citep[e.g.,][]{platzman75,Egbert2004, green10}. The resonant period of the ocean basins on Venus (and Earth) is measured in hours and the tidal period in our Venusian simulation is measured in tens of days, so the basins cannot be near resonance. For example, the Venus basin in the southwest centered at (45$^\circ$S, 20$^\circ$W) is approximately 2500 km across. The propagation speed, $c_g = \sqrt{gH}$, of the tidal wave would be about 95 $\mathrm{m\ s^{-1}}$ if the basin is 1000 m deep. A half-wavelength resonance, i.e., a 5000 km long wave, in that basin would require a tidal period of 14.6 hours. Because the simulations with a faster rotation will be closer to this number we expect the tides to get more energetic as the tidal period decreases.

\subsection{Sensitivity to rotation rates}
\begin{figure}
  \centering
  \includegraphics[width=0.75\textwidth,trim={0 10cm 0 7cm},clip]{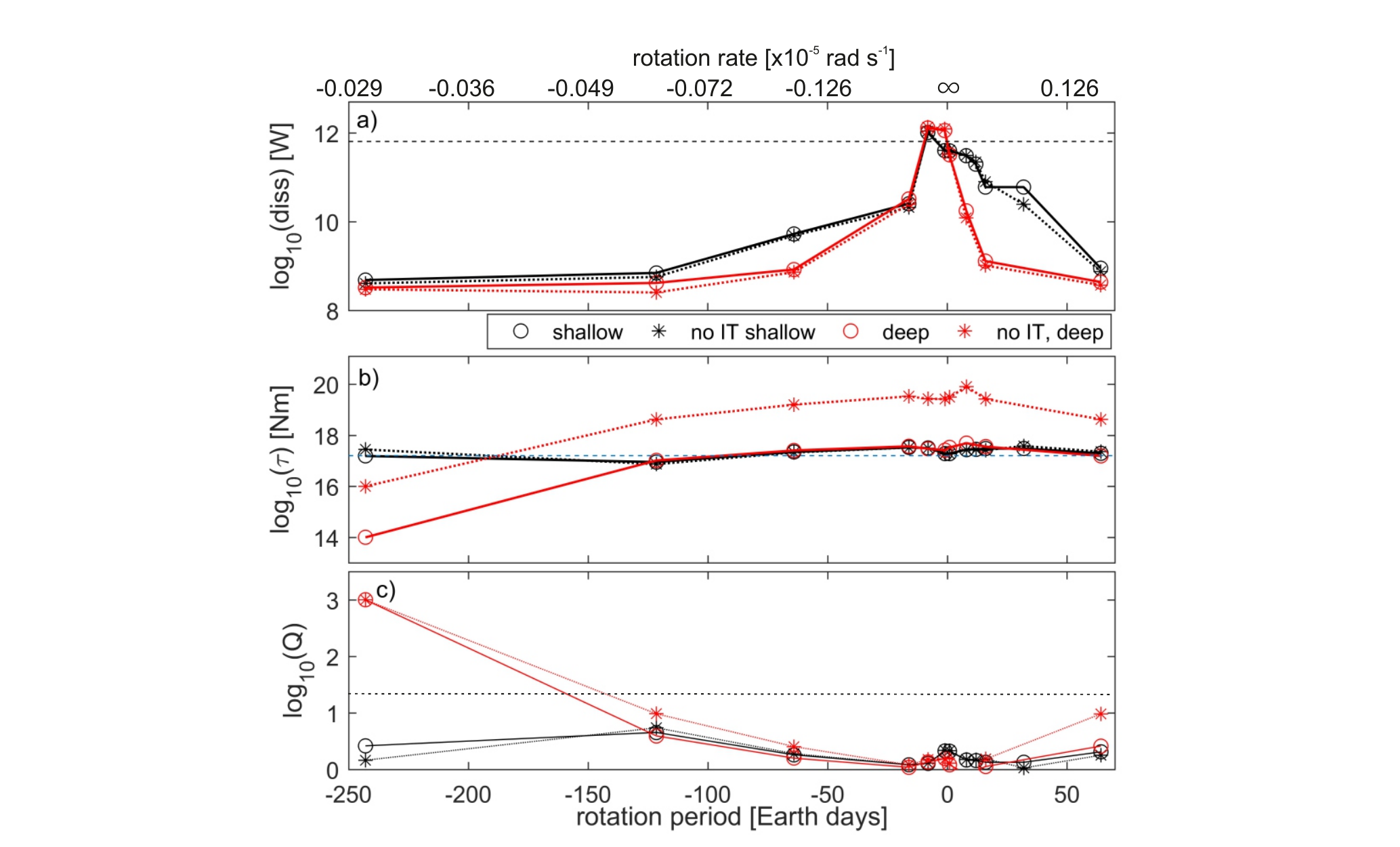}
  \caption{a) Horizontally integrated dissipation rates for the different simulations plotted against the rotation rate (black: shallow, red: deep simulations; circles: with conversion, stars: without conversion/no IT). Note that the y-axis is logarithmic. The dashed horizontal line is the corresponding dissipation rate in Earth's present day ocean. \newline
  b) As in a) but showing the torque.
  \newline
  c) As in a) but for the tidal damping factor, $Q$ (i.e., the number of cycles to obtain an e-folding decay of the amplitude.}\label{fig:DvsR}
\end{figure}
Indeed, an altered rotation rate does change the picture dramatically. As an example, we show the results from the shallow prograde 8-day simulation in Figs.~\ref{fig:bathy}d and f, where a more energetic tide would be generated compared to a shallow present day Venusian ocean. The associated globally integrated dissipation rate is now more than three orders of magnitude larger than under present day conditions because there are regions between the continents where the tide can be amplified due to (near-)resonance (Fig.~\ref{fig:bathy}d and f).

This phenomenon is further highlighted in Fig.~\ref{fig:DvsR}a, which shows the horizontally integrated dissipation rates from all the simulations. It is clear that the dissipation is dependent on the rotation rate, with a maximum in the dissipation at -8 days and slow decline until 32 days. Interestingly, the deep bathymetry simulations have a sharp peak in dissipation at the -8 to -1 day periods, suggesting that at the lower rotations rates, conversion is more effective at dampening the tides. We conclude that a low rotation rate, with periods of several tens of days, will only support a weak tide, regardless of rotational direction.

The results in Fig.~\ref{fig:DvsR}a also show the robustness in terms of stratification: the average ratio of the integrated rates between runs with and without tidal conversion (noIT) is a factor of 2, whereas the ITx10-simulation changed the dissipation by a factor of 6, to 0.8GW for the shallow bathymetry. Similar results were found for the other three ITx10-cases, where the extremely strong stratification increased the dissipation rate with a factor 3--10 (not shown). Our results for a shallow Venusian ocean may thus represent an underestimate if ancient Venus was very strongly stratified, or a slight overestimate if it was vertically well mixed. This robustness has been reported on Earth before under less extreme circumstances by \citet{Egbert2004} and \citet{Green2013a} and gives confidence in our conclusions. Because of the uncertainty in the stratification on ancient Venus we opt to continue our focus on the Venus-shallow case until additional information is available.

Furthermore, a sensitivity simulation with Earth's bathymetry on Present Day Venus shows a dissipation rate some 40\% larger than from the Venusian shallow simulation (0.13 vs. 0.18 GW; not shown), suggesting that the rotation rate exhibits a first order control on the dynamics of tides. We also performed a few simulations with Venus' bathymetry and a 4300 m deep ocean and one with an 80 m deep ocean (not shown; denoted very deep and very shallow in the following). The results from the very deep (4300 m) simulation are less energetic than the deep, whereas the very shallow simulation becomes slightly more energetic than the shallow simulations. For the Present Day Venus rotation, the four depths -- from very shallow to very deep -- span approximately 4 orders of magnitude in dissipation and thus provide a sensitivity range of potential dissipation rates in an ocean on an Earth-like planet.

\subsection{Consequences}
%
\begin{sidewaystable}[t!]
  \centering
  \caption{A summary of the results from the shallow (mean depth = 330 m) and the deep (mean depth = 830 m) simulations. $d\Omega/dt$, where $\Omega$ is the rotation rate, is computed from Eq.~(\ref{eq:dOdt}) and shown as $10^{-9}$ per 1 Myr. $D$ is the tidal dissipation rate shown in Fig.~\ref{fig:DvsR}a, and $Q$ is tidal quality factor, showing the e-folding time scale in terms of tidal periods.}\label{tab:res}
\begin{tabular}{ll|lll|lll}
\hline
 & & & with IT & & & noIT &  \\
 Daylength & Mean depth & $d\Omega/dt$ & $D$  & $Q$ & $d\Omega/dt$ & $D$  & $Q$  \\
  days  & m  & $\times 10^{-9}\ \mathrm{Myr^{-1}}$ & GW & & $\times 10^{-9}\ \mathrm{Myr^{-1}}$ & GW &  \\
\hline
 -243  & 330  & 68.41  & 0.15 & 2.6 & 122.83 & 0.07 & 1.5  \\
 -121.5  & 330  & 39.67  & 0.32 & 4.5 & 32.78  & 0.16 & 5.5  \\
 -64 & 330  & 98.62  & 4.31 & 1.8 & 94.18  & 3.71 & 1.9  \\
 -16 & 330  & 147.70 & 18.19  & 1.2 & 146.71 & 11.09  & 1.2  \\
 -8  & 330  & 139.87 & 652.64 & 1.3 & 138.95 & 658.75 & 1.3  \\
 -1  & 330  & 84.61  & 88.01  & 2.1 & 82.85  & 86.15  & 2.2  \\
 0.9972  & 330  & 85.62  & 86.12  & 2.1 & 83.47  & 84.05  & 2.1  \\
 8.02  & 330  & 122.41 & 228.06 & 1.5 & 119.48 & 212.45 & 1.5  \\
 12  & 330  & 124.12 & 155.90 & 1.4 & 119.95 & 159.14 & 1.5  \\
 16.04 & 330  & 133.60 & 44.34  & 1.3 & 116.10 & 46.89  & 1.5  \\
 32  & 330  & 133.60 & 44.34  & 1.3 & 168.22 & 19.43  & 1.1  \\
 64.15 & 330  & 87.48  & 0.44 & 2.0 & 100.27 & 0.28 & 1.8  \\
 -243  & 830  & 987.08 & 0.03 & 1.0 & 12.61  & 0.00 & 13.5 \\
 -121.5  & 830  & 45.62  & 0.04 & 3.9 & 18.54  & -0.01  & 9.7  \\
 -64 & 830  & 113.06 & 0.17 & 1.6 & 70.58  & 0.02 & 2.5  \\
 -16 & 830  & 163.91 & 24.33  & 1.1 & 148.92 & 10.15  & 1.2  \\
 -8  & 830  & 131.54 & 777.19 & 1.4 & 118.63 & 632.76 & 1.5  \\
 -1  & 830  & 115.47 & 307.53 & 1.6 & 116.44 & 313.28 & 1.5  \\
 0.9972  & 830  & 148.46 & 239.60 & 1.2 & 138.04 & 186.39 & 1.3  \\
 8.02  & 830  & 216.23 & 10.98  & 1.0 & 360.37 & 3.13 & 1.0  \\
 16.04 & 830  & 158.82 & 0.44 & 1.1 & 117.23 & 0.09 & 1.5  \\
 64.15 & 830  & 68.93  & 0.06 & 2.6 & 18.62  & 0.01 & 9.6\\
 \hline				
 \end{tabular}
\end{sidewaystable}
%
Fig.~\ref{fig:DvsR}b shows the associated tidal torque, computed from Eq.~(\ref{eq:torque}) and the dissipation rates in Fig.~\ref{fig:DvsR}a. Using this torque in Eq.~(\ref{eq:dOdt}) shows that the dissipation in the Venus-shallow simulation could change the rotation rate of present day Venus by over $6.8\times 10^{-8}\ \mathrm{rad~Myr^{-1}}$. This is equivalent to a day-length change of nearly 72 days per million years (Table~\ref{tab:res}), or equal to about half of the observed change in day-length on present day Venus of about 7 minutes over the past 40 years \citep{Mueller2012a, Navarro2018}, which has been attributed to the present day torque exerted by the dense atmosphere. We have thus shown a similar magnitude effect should ancient Venus have had an ocean and its modern rotation rate. For the other simulations, the changes in day length are less extreme, even though the dissipation rates are higher: the 8-day simulation shown in Fig.~\ref{fig:bathy}e--f would induce a change of $12.2\times 10^{-8}\ \mathrm{rad\ s^{-1}}$, or 2.6 hours per million years (or 0.35 s per 40 years), whereas the slower prograde simulations, with day-lengths of 32 or 64 days, show a change of 8.7 to 13.3$\times 10^{-8}\ \mathrm{rad\ s^{-1}}$ (up to 4.5 days per million years). For comparison, the present day rate of change on Earth is 20 seconds per million years, or $1.7\times 10^{-8}\ \mathrm{rad\ s^{-1}}$.

These results suggest that a faster-spinning ancient Venus with an ocean would have slowed rapidly due to the tidal torque. Even a potential short-lived ocean could have slowed the rotation rate by several days, especially if Venus' rotation rate was initially slower than 1 Earth day.

\section{Summary}
Our aim here is not explicitly to simulate tides on ancient Venus, but rather to provide a sensitivity study of a plausible range of tidal dissipation rates and the associated effects should Venus have had an ocean. The results show that even a short-lived ocean on a faster-spinning Venus had the potential to host a solar tide with amplitudes exceeding 0.5m. While weak compared to present day Earth, the substantial torque set up by the tide had the potential to slow down Venus' rotation rate by up to 5 days per million years for a faster spinning planet. Venusian tides may thus have had a profound impact on the rotational evolution of Venus.

We have done simulations using tidal conversion based on present day Earth, which is a coarse approximation. Would the ocean used here be stratified under the conditions described by \citet{Way2016}? This is an intriguing question from an oceanographic perspective, but left for future studies. The tests with the conversion coefficient reduced or increased by an order of magnitude did not result in changes in dissipation of an order of magnitude but a factor of about 2. We also know that conversion is a crucial energy source in Earth's ocean, and including both the simulations with and without conversion acts as another sensitivity study.

The results point to a fundamental aspect of planetary tidal dynamics: the influence of day length on the tidal amplitudes. To first order, tidal dissipation is set by the planet's continental configuration \citep{Green2018}. Shelf-sea extent and sea-level then becomes important in basins that are near-resonant. Here we argue that, to zeroth order, tidal dissipation is set by the planet's rotation rate: the slower the rotation the weaker the tides. \textit{In extremis}, a tidally locked planet will not have any tidal dissipation induced by the locked body, and will have an infinitely slow rotation rate in relation to it.

\acknowledgments
JAMG received funding from NERC, grant NE/I030224/1. RB acknowledges support from NASA grant NNX15AN35G. This work was supported by the NASA Astrobiology Program through collaborations arising from our participation in the Nexus for Exoplanet System Science, and by the NASA Planetary Atmospheres Program. MJW is thankful for support from the Goddard Space Flight Center's Sellers Exoplanet Environments Collaboration (SEEC), which is funded by the NASA Planetary Science Division's Internal Scientist Funding Model. Simulations were done using Supercomputing Wales and their support is greatly appreciated. Comments from one anonymous reviewer vastly improved the manuscript.


%



\end{document}